# GPS-Spoofing Attack Detection Mechanism for UAV Swarms

Pavlo Mykytyn[1, 2], Marcin Brzozowski[1], Zoya Dyka[1,2] and Peter Langendoerfer[1, 2]
[1] IHP - Leibniz-Institut für innovative Mikroelektronik, Frankfurt (Oder), Germany
[2] BTU Cottbus-Senftenberg Cottbus, Germany
pavlo.mykytyn @b-tu.de

*Abstract* — **Recently autonomous and semi-autonomous Unmanned Aerial Vehicle (UAV) swarms started to receive a lot of research interest and demand from various civil application fields. However, for a successful mission execution UAV swarms require Global Navigation Satellite System (GNSS) signals and in particular Global Positioning System (GPS) signals for navigation. Unfortunately, civil GPS signals are unencrypted and unauthenticated, which facilitates the execution of GPS spoofing attacks. During these attacks adversaries mimic the authentic GPS signal and broadcast it to the targeted UAV in order to change its course, force it to land or crash. In this study, we propose a GPS spoofing detection mechanism capable of detecting single-transmitter and multi-transmitter GPS spoofing attacks to prevent the above mentioned outcomes. Our detection mechanism is based on comparing the distance between each two swarm members calculated from their GPS coordinates to the distance acquired from Impulse Radio Ultra-Wideband ranging between the same swarm members. If the difference in distances is larger than a chosen threshold the GPS spoofing attack is declared detected.**

## I. Introduction

The use of Unmanned Aerial Vehicles (UAVs) also known as drones for private, commercial, and military [1] applications has increased drastically in the past few years [2]. The use cases for UAVs include search and rescue [3], agricultural field management [4], construction and mining [5, 6], and film industry [7]. UAVs are easily adaptable to a custom mission, require little maintenance and have low operational costs. However, execution of an autonomous mission by a UAV relies heavily on the use of the Global Navigation Satellite System (GNSS) for positioning, navigation, and mission planning. The global navigation satellite system, specifically Global Positioning System (GPS), is the primary location estimation technology used by UAVs due to its global coverage and accuracy. GPS is based on a constellation of a total of up to 31 operational satellites, from which up to 3 can be in reserve or standby mode. Each GPS satellite transmits two signals, one civilian unencrypted and one military encrypted and authenticated. The civilian GPS signal was never meant for critical or security applications, despite the fact that nowadays, this is exactly what it is often used for [8]. Even though UAVs have received massive improvements in terms of design, hardware and software in the past ten years [9], many security issues and concerns remained largely untouched. Due to the broad field of applications, civilian GPS signals are not secured. Only military GPS signals are encrypted and authenticated, but they are generally unavailable for civilian applications. A substantial issue with the GNSS, and GPS in particular, is that the received signals are extremely weak. Due to the high path loss between the satellite and earth, the received surface signal power is only equal to about ≈ -130 dBm [10]. This is around 20 dB lower than the thermal noise floor, assuming the signal power is evenly spread in a 2-MHz bandwidth. This particularly low signal power makes the technology exceptionally vulnerable to interferences, deliberate jamming, and spoofing attacks. The consequences of a GPS spoofing attack can be quite harmful, e.g., change of the flying course, sudden acceleration or decelerations, collision with other UAVs, or forced switch into the manual or landing mode by transmitting the no-fly zone coordinates. To effectively mitigate GPS spoofing attacks we need a reliable spoofing attack detection mechanism. In the past years, multiple detection techniques against a GPS spoofing attack have been proposed [11]. However, many of them require modifications of the receiver's antennas, installation of additional GPS receivers, or access to the physical properties of the GNSS signal, such as signal strength level and its arrival direction. Thus, in most cases it requires modifications of existing or integration of the additional hardware components.

In this study we propose a GPS spoofing attack detection mechanism for UAV swarms presented on an example of an IR-UWB assisted UAV swarm. The detection mechanism takes advantage of the GPS localization data reported to the ground station by each UAV. The localization results of two UAVs in a swarm are then checked for equality. If the reported coordinates are identical, it indicates that the GPS signal received by the drones is spoofed. To tackle the case when only one of the two UAVs is affected by the spoofed GPS signal, we additionally compare the distance between them from their GPS data to the distance from IR-UWB ranging, or other similar distance ranging technologies. Thus, not limiting our detection mechanism to one distance ranging technology only.

The remainder of this paper is structured as follows. In Section II we describe the GPS spoofing attacks on UAV swarms. Section III provides an overview of the related work. The proposed GPS spoofing attack detection mechanism is presented in Section IV. Section V presents future work directions and concludes the paper.

## II. GPS SPOOFING ATTACKS

GPS spoofing and jamming attacks are among the most common cyber-attacks on UAVs registered so far as reported in [12]. GPS Spoofing and jamming are often executed simultaneously. The adversary starts jamming the authentic GPS signal to force a UAV into the lost satellite signal state. At the same time, the adversary initiates transmission of the imitated GNSS signal with a much higher signal power using a GNSS spoofing device located nearby. The threat of a GPS spoofing has significantly increased with the arrival of cheap Software-defined radios (SDRs) to the market. SDRs can easily imitate the authentic GPS signal and cause a serious deviation in the UAV's intended flight path. The effects of a GPS spoofing attack could be devastating, including a collision, crash or even theft of the UAV. This poses a significant safety threat not only to UAVs, but also to remote pilots and any people around.

The position estimation of a GPS receiver is based on a time of arrival of the received satelite signals from four or more satelites orbiting around Earth [13]. By using the signal's time of arrival from each of the satellites, knowledge about their position in space, and multilateration, the GPS receiver can estimate its position on Earth and receive local time. A GPS spoofing attack, on the other hand, is an intentional mimicking of the legitimate GNSS signal in order to achieve a malicious objective. If the GPS spoofing attack is not detected, its consequences could be potentially even more harmful than the consequences of a GPS jamming attack. As long as the GPS spoofing remains undetected by the targeted UAV, its navigation remains under control of the attacker. Execution of the sophisticated GPS spoofing attacks has become less expensive and more feasible due to the advancement of the SDR technology.

Two classical GPS spoofing attack scenarios are *single-transmitter attack* and *multi-transmitter attack*.

***Single – transmitter attack***

In this attack scenario, the attacker uses a single transmitter to initiate the spoofing attack. Thus, all of the spoofing signals are identical and are coming from the same source. Overpowering the authentic GPS signal with the spoofed one is the most common way to execute this attack. However, depending on the technical capabilities of the attacker, an additional GPS jamming attack can take place before the execution of the spoofing attack. Fig.1 represents the single–transmitter GPS spoofing attack on a swarm consisting of 3 UAV.

Depending on the malicious objective, the attacker can use a Software-defined radio tuned to the GPS frequency (1575.42 MHz) to generate its own spoofed GPS signal or simply replay the authentic GNSS signal recorded prior to the attack.

The number of drones affected by the GPS spoofing attack might vary depending on the distance from the attacker to the swarm, as well as distance intervals between the UAVs within a swarm. UAVs located in the affected area receive identical spoofed GPS signals and compute nearly the same localization result with a minor time difference depending on the distance between the attacker and a particular UAV [13].

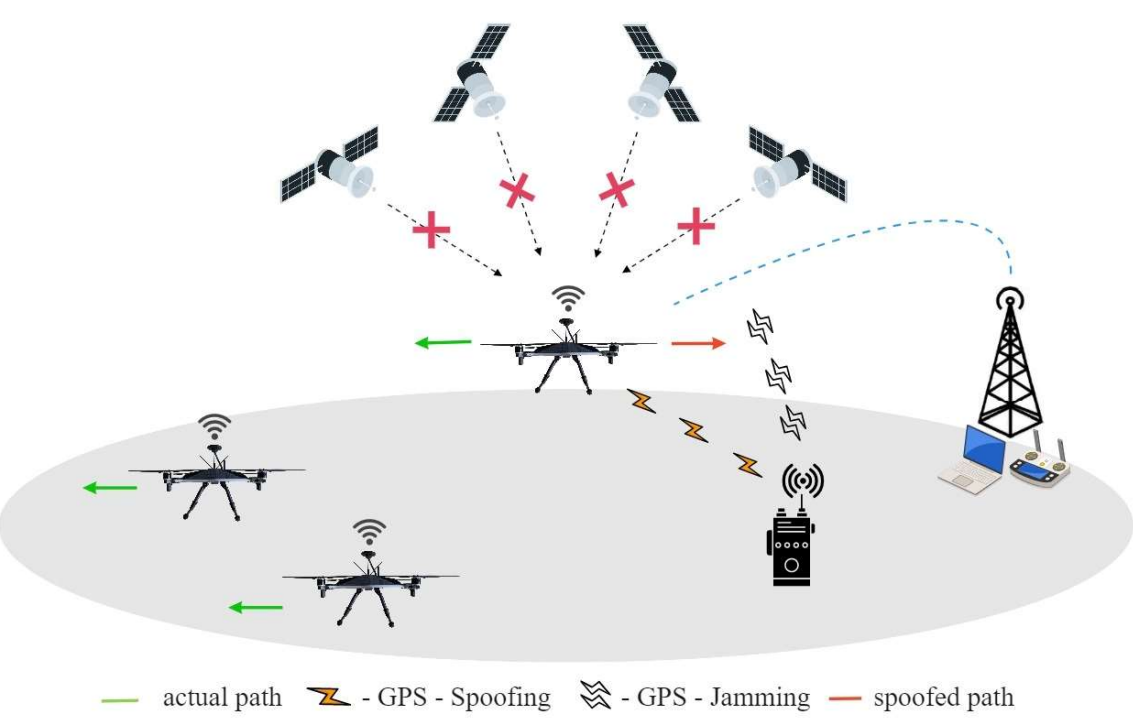

Fig.1: Single-transmitter GPS spoofing attack

***Multi – transmitter attack*:**

The attacker uses two or more transmitters to initiate the spoofing attack. Thus, the spoofing signals can be coming from different sources to cover a larger area or send different spoofing signals from each of the transmitters. In this attack scenario, the adversary utilizes multiple transmitters to cover larger areas or to transmmit diverse spoofing signals to different UAVs placing the transmitters in various locations. Fig.2 represents the multi–transmitter GPS spoofing attack on UAVs using two spoofing devices.

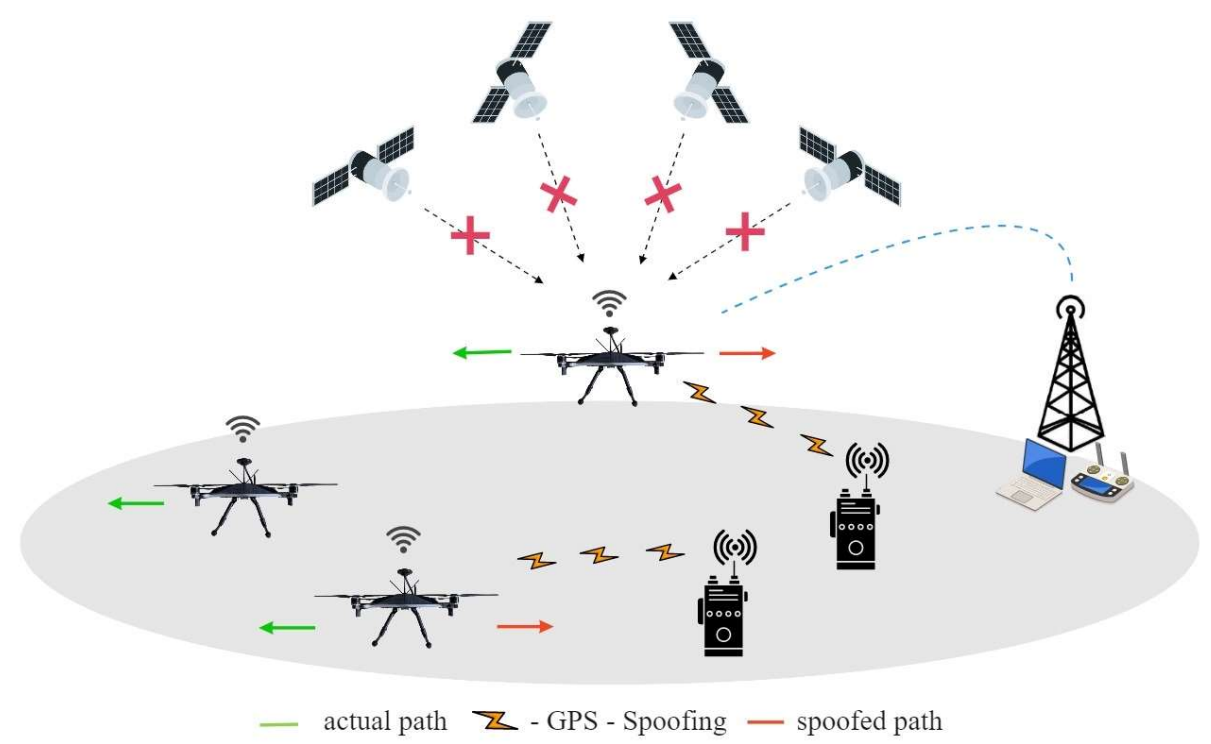

Fig.2: Multi-transmitter GPS spoofing attack

Depending on a distance between the UAVs and positions of the spoofing signal transmitters, the affected UAVs can either receive the same or different spoofed GPS signals. In this scenario, an attacker can force two or more UAVs to move on a collission course, accelerate, switch into the manual mode, or land by transmitting the no-fly zone coordinates. If the spoofing attack is not detected in time, it can become a significant safety hazard for the UAVs and people around it.

## III. Related Work

The vulnerability of GPS receivers to spoofing is not a new topic and has been discussed in the literature previously. Several studies have reviewed the GPS spoofing detection techniques [11, 14], however, only very few of them do not require any modifications in the hardware, installment of additional antennas and receivers, or access to the physical parameters of a GPS signal.

Table 1 presents a few distinct approaches to the GPS spoofing detection available in the literature, their drawbacks, and detection rates. E.g. some authors combine the GPS localization with other sensor measurements such as internal measurement unit (IMU), gyroscope or accelerometer [15, 16] to detect the anomalies and deviations in GPS localization. Some authors suggest using multiple receivers with a fixed distance between them [13] or antenna arrays to estimate the direction of arrival of the spoofed GPS signal and detect the spoofing attack [17, 18]. Other authors rely on the GPS signal's time of arrival correlations [19, 20], video stream analysis from the UAV's onboard cameras [21], or the use of cellular towers and cellular network infrastructure to aid the detection process [22, 23]. The detection mechanisms presented in the table are sorted by their corresponding detection rate percentages.

TABLE I. An overview of the GPS spoofing detection mechanisms

| Mechanism | Approach | Limitations and Drawbacks | Detection rate |
|---|---|---|---|
| Vehicular Network (VN) based [19] | time of arrival correlations | Only applicable to a VN; requires vehicle to vehicle communication | n/a |
| Camera's video stream based [21] | Correlation between frames and coordinates | Speed and terrain dependable; requires camera to be pointed downward at all times | n/a |
| Crowd-GPS-Sec [20] | time of arrival, signal properties and ADS-B message analysis | Additional ground-infrastructure needed | 75 % |
| Cellular network based [23] | Position validity cross-check | Requires a cellular module; requires cellular network coverage | 95 % |
| Gyroscope and Accelerometer based [15] | Position estimation | Requires additional motion sensors for accurate detection. | 96 % |
| IMU based [16] | Machine learning | Computational overhead; limited by the initial training set of data. | 96.3 % |
| Multi-receiver [13] | Position comparison | Requires additional hardware; not suitable for small UAVs | 99 % |
| Antenna Array [17] | GPS signal's direction of arrival | Requires additional hardware; not suitable for small UAVs | 99 % |
| Cellular network based [22] | Position verification | Requires additional cellular modules; not suitable for rural areas | 100 % |

In [19] authors proposed a decentralized spoofing detection mechanism by utilizing the help of short-range communication technologies. Vehicles exchange their measured GPS code pseudo-range measurements between themselves and then perform linear operations to derive independent statistics related to each neighboring vehicle. Derived statistics are then reported to the head-vehicle to perform spoofing detection. The downside of this approach is that it requires a specific hierarchy in the vehicular network, as well as a vehicle-to-vehicle communication. The processing part takes place onboard each vehicle as well, introducing a computational overhead. The authors mention that the probability of false alarms in close to zero, however, they do not provide the experimental or simulated detection rate statistics in their work.

In [21] the authors suggest a spoofing detection mechanism based on analyzing UAV's camera video stream. The proposed mechanism collects frames from the video stream and their GPS coordinates. The correlation between the drone's movements from the GPS coordinates and real-time video stream frames is calculated to determine if a spoofing attack takes place. The downside of this method is that its performance highly depends on the terrain, speed, altitude and ambient light. Additionally all of the processing from the UAV's camera takes place onboard the UAV, thus introducing computational overhead. The authors claim that the method can detect any GPS spoofing attack. However, they do not provide experimental or simulated detection rate statistics in their work.

In [20] the authors proposed a spoofing detection method called Crowd–GPS–Sec, that leverages crowd-sourcing to monitor the automatic dependent surveillance–broadcast (ADS-B) traffic of an aircraft. ADS-B is responsible for the GPS-derived position advertisements of an aircraft. UAVs or other aircrafts periodically broadcast ADS-B messages for air traffic control purposes. Spoofing attacks are detected and localized by an independent infrastructure on the ground which continuously analyzes the contents and the time of arrival of these messages. The downside of this method is the dependability on the third party ground infrastructure for the detection. The authors report that this mechanism can detect up to 75% of all of the GPS spoofing attacks based on their simulation.

In [23] the authors suggested a spoofing detection method based on utilizing the received signal strength (RSS) of the periodically transmitted telemetry messages from a UAV to the base station, which include the UAV's GPS coordinates. Additionally, with the help of 5G modules, they estimate the UAV's position based on the time of arrival , angle of arrival and RSSI from the surrounding 5G base stations to map an adaptive trustable residence area (ATRA) for a UAV. If the GPS coordinates received from the UAV locate it out of the ATRA, the UAV's GPS position is declared spoofed, otherwise it is considered authentic. The downside of this method is that it relies on the 5G cellular coverage for the ATRA estimation. The authors reported a 95% detection rate in urban and suburban areas.

In [15] the authors proposed a method based on combining the angular velocity and linear acceleration measurements of a gyroscope and an accelerometer to estimate the UAV's position and compare it with the received GPS coordinates. The downside of this approach is that despite its high detection rate, reported to be at 96%, it requires intensive computation in order to be able to keep track of the UAV movements in a 3D space and compare it with the reported GPS positions.

In [16] the authors presented a two-step detection mechanism, they utilize the IMU measurements such as angular velocity, acceleration, direction of travel and the change in received GPS coordinates from existing UAV flight logs to train the machine learning model using Genetic Algorithm and distributed gradient-boosted (XGBoost) decision tree machine learning library. Once the model is trained they deploy it on a UAV and fine tune it for better detection. Similar to the previous mechanism, it requires significant computational capabilities in order to process the IMU measurements and operate the ML algorithm, which makes it unsuitable for smaller UAVs.

Jansen et al. [13] proposed a multi-receiver approach for the spoofing detection. The detection mechanism works by comparing the fixed distance between the receivers to the distance between their reported locations. In case of a GPS spoofing attack, the reported location difference from both of the receivers will be very close to zero. Otherwise, if the signal is authentic the reported distance will be close to the actual fixed distance between the receivers. Despite its high 99% detection rate, the biggest downside of this mechanism is that it requires at least two GPS receivers with a fixed distance between them, which for a relatively small UAV is unfeasible.

In [17] the authors presented a spoofing detection mechanism based on multiple receiver antennas. It requires a specific array arrangement of receiver antennas to estimate the direction of arrival and compare the phase delays of a GPS signal. This mechanism relies on the fact that an attacker usually transmits several GPS signal codes from the same antenna while the authentic signals are transmitted by different satellites from various directions. The downside of this approach is that even with its high detection rate at 99%, it requires sophisticated signal processing algorithms and multiple external antennas to be installed on a UAV in order to facilitate the GPS spoofing attack detection process.

In [22] authors present a detection mechanism that takes advantage of the cellular infrastructure to compute an approximate location of the moving vehicle based on cell tower beaconing messages and verify that the GPS coordinates received by the UAV are within this approximate location. Similar to the mechanism presented in [23], its biggest downside is that even with its perfect detection rate claimed by the authors to be at 100 %, it requires additional cellular modules and is totally reliant on cellular coverage in the area of operation.

Each of the above mentioned techniques and detection mechanisms has its own disadvantages. Some require sophisticated signal processing [17, 23] or intensive computational resources [15, 16] to run machine learning algorithms, which are not feasible on smaller UAVs . Others demand additional hardware components [13, 17, 20] to be installed on the ground or onboard a UAV. Smaller size commercial and civilian UAVs usually rely on microcontrollers with limited computational capabilities and space onboard for supplementary hardware components, which renders the above mentioned techniques inapplicable. Therefore, we suggest an easily executable spoofing detection mechanism suitable for small and moderate size UAV swarms that does not require sophisticated signal processing or any additional ground based infrastructure and can be implemented using any distance ranging technology similar to the IR-UWB ranging.

## IV. PROPOSED SPOOFING DETECTION MECHANISM

In this section, we present our GPS spoofing attack detection mechanism on an example of IR-UWB assisted UAV swarms. Our method relies on the plausibility confirmation of GPS localization data reported to the ground control station (GCS) by each drone in the swarm. Let's assume that our UAV swarm consists of *n* UAVs, further denoted as $UAV_1 \ldots UAV_n$. From the GPS localization data received by the GCS we calculate the distance between the $UAV_1$ and the rest of the drones in the swarm ($UAV_2 \ldots UAV_n$). Additionally, GCS also requests the IR-UWB distance ranging data between the $UAV_1$ and the rest of the swarm ( $UAV_2$ … $UAV_n$ ). The IR-UWB distance measurements between the UAVs in a swarm are taken regularly for collision avoidance purposes anyway, thus, they would not introduce an additional energy constraint. Therefore, by comparing the two distances we can determine whether the reported GPS coordinates are authentic or spoofed. UWB distance ranging has many advantages in terms of precision, energy consumption, multipath robustness, and resistance to interference and jamming, thus making it a reliable source of distance measurements for the confirmation of the GPS data plausibility. IR-UWB ranging technology is based on short, low energy, narrow pulses over a wide spectrum of frequencies with a high time resolution. It provides the ability to measure the distance between two objects with a precision of a few centimeters and an operational range of up to 150 meters [24] as claimed by one of the leading manufacturers. However, our spoofing detection mechanism can be coupled with other similar distance ranging technologies as well.

### *A. Detection mechanism*

We differentiate between two attack scenarios mentioned above, namely a Single-transmitter and a Multi-transmitter GPS spoofing attack. We utilize different detection factors for each spoofing attack scenario. In case of a single - transmitter attack, there are two possible outcomes of the GPS spoofing attack, namely only one affected UAV or two and more affected UAVs. Below we describe in detail the detection process for each of the outcomes.

***Case 1***: only the $UAV_1$ is affected by the spoofed GPS signal in the single transmitter scenario.

The reported GPS coordinates to the GCS by the $UAV_1$ consist of the latitude (lat) and longitude (lon) expressed in degrees that are further denoted as $GPS\ (UAV_1) = (lat_1, lon_1)$. The distance between the spoofed $UAV_1$ and the other – not spoofed – vehicle $UAV_i$, where $i=2,3,\ldots, n$ is calculated from their reported GPS coordinates by applying the "Spherical law of Cosines" shown in the formula (1) below, and is further denoted as $d_{GPS}(UAV_1, UAV_i)$.

To account for the altitude difference between the $UAV_1$ and $UAV_i$, in case of them flying at different altitudes, the distance $d_{GPS}$ must be adjusted by utilizing the Euclidean distance calculation between two coordinates in a three-dimensional space. In our case, adjustment for altitude can be calculated using the Pythagorean Theorem where $d_{GPS}$ serves as one side of the right triangle and altitude difference as the other. Thus, the hypotenuse of that right triangle is the real-world distance between the UAVs with an account to their respective altitude.

The acquired IR-UWB ranging distance between the spoofed $UAV_1$ and other – not spoofed – $UAV_i$ is further denoted as $d_{UWB}(UAV_1, UAV_i)$.

Formula (1) is the realization of the "Spherical law of Cosines" where $\boldsymbol{\varphi}$ is the latitude and $\boldsymbol{\lambda}$ is the longitude, expressed in radians, $\Delta\boldsymbol{\lambda}$ is the longitude difference between the $UAV_1$ and the $UAV_i$, and is equal to $\Delta\boldsymbol{\lambda} = (\lambda_i - \lambda_1)$. $\boldsymbol{R}$ is the Earth's mean radius that is equal to 6,371km.

$$d_{GPS}(UAV_1, UAV_i) = \arccos\ (\sin\varphi_1 \cdot \sin\ \varphi_i + \cos\ \varphi_1 \cdot \cos\ \varphi_i \cdot \cos\Delta\lambda) \cdot R \qquad (1)$$

In this scenario, after calculating the $d_{GPS}(UAV_1,UAV_i)$, we can compare it to the $d_{UWB}(UAV_1, UAV_i)$. If the comparison result reveals a large disparity between the two distances, that will be an indication of a GPS spoofing attack.

Considering that in most cases, each GPS receiver's localization error is not larger than 4.9 meters [25] and the IR-UWB ranging error lays within 30 cm. [26], the difference in measured distances should not exceed the sum of these two errors. In fact, if the difference between the two compared distances is larger than the measurement error threshold (denoted further as $d_{THR}$), it indicates that either one or the other UAV is under a GPS spoofing attack. The $d_{THR}$ can be individually adjusted for each UAV swarm depending on the terrain and quality of the received GPS signal to improve its accuracy and effectiveness. Additionally for UAV swarms with an average operational distance of less than 10 meters between the swarm members, so called differential GPS (DGPS) system or Real-time Kinematic positioning (RTK) is frequently used. They provide positioning accuracy of 1-3 centimeters [27]. In this case the $d_{THR}$ for GPS spoofing attack detection can be reduced significantly.

***Case 2***: two or more UAVs are affected by the GPS spoofing attack in the single transmitter scenario.

In this case, two or more UAVs – for example $UAV_i$ and $UAV_j$ – receive the same spoofed signal, i.e. their calculated GPS localization results would be almost identical with a minor time difference depending on the signal's time of arrival from the spoofing device. Thus, the distance between the affected UAVs calculated from their GPS coordinates will be very close or equal to zero, i.e. $d_{GPS} \approx 0$. This fact alone can be utilized for a successful GPS spoofing attack detection. For the case when the minimal distance between two UAVs in a swarm is controlled automatically by a collision avoidance algorithm, the presence of a single-transmitter GPS spoofing attack can be easily detected just by comparing the GPS localization results of the UAVs, without needing to compare them to the IR-UWB ranging results, thus not requiring any additional ranging hardware onboard. However, by comparing $d_{GPS}$ to $d_{UWB}$ between each two UAVs – $UAV_i$ and $UAV_j$ – we can reduce the false positives and make the detection mechnism more reliable. It is important to point out that $d_{UWB}$ cannot be equal or close to zero, otherwise it would mean that the $UAV_i$ has collided and crashed into the $UAV_j$.

***Case 3***: two or more UAVs are affected by the GPS spoofing attack in the multiple-transmitter scenario.

In this case we can either have the same spoofed GPS signal broadcasted by multiple transmitters to cover larger areas, or have different GPS spoofing signals broadcasted by each of the transmitters. In the first outcome, the GPS spoofing attack would be identical to the Single-transmitter GPS spoofing attack, but with a broader affected area. In the second outcome, each of the two UAVs from the swarm might receive diverse spoofed GPS signals and thus, compute distinct localization results. However, in this case we can still apply our detection mechanism and compare the distance calculated from the GPS coordinates to the UWB ranging result to reveal the inconsistencies and thus, detect the presence of a GPS spoofing attack.

Table II represents both single-transmitter and multi-transmitter GPS spoofing attack scenarios including four outcomes (two from each of the scenarios), as well as all the detection criteria used by our detection mechanism.

Ideally, each UAV swarm should have an integrated collision avoidance system. If that is the case, IR-UWB ranging or similar measurements between the swarm members have to be taken regularly and sent to the GCS in order to avoid collisions with other swarm members as shown on the Fig.3. Thus, in this case, the effect of our detection mechanism on the overall UAV's energy consumption, performance, and battery life would be extremely insignificant. However, if the collision avoidance system is not integrated, IR-UWB measurements would need to be taken additionally in order for the spoofing detection mechanism to work properly, which could increase the power consumption and decrease battery life.

TABLE II. GPS SPOOFING ATTACK DETECTION CRITERIA

| **Spoofing attack Scenario** | **Affected UAVs** | **Attack detection criteria** |
|---|---|---|
| Single-transmitter | 1 | $\begin{cases} GPS(UAV_1) \neq GPS(UAV_i), i \in \{2, \ldots, n\} \\ \lvert d_{GPS}^{1,i} - d_{UWB}^{1,i} \rvert > d_{THR} \end{cases}$ |
| | >1 | $GPS(UAV_1) \cong GPS(UAV_i), i \in \{2, \ldots, n\}$ or $\lvert d_{GPS}^{1,i} - d_{UWB}^{1,i} \rvert > d_{THR}$ |

| Multi-transmitter | 1 | $\begin{cases} GPS(UAV_1) \neq GPS(UAV_i), i \in \{2, \dots, n\} \\ \lvert d_{GPS}^{1,i} - d_{UWB}^{1,i} \rvert > d_{THR} \end{cases}$ |
| --- | --- | --- |
| | >1 | $GPS(UAV_1) \cong GPS(UAV_i), i \in \{2, \dots, n\}$ or $\begin{cases} GPS(UAV_1) \neq GPS(UAV_i), i \in \{2, \dots, n\} \\ \lvert d_{GPS}^{1,i} - d_{UWB}^{1,i} \rvert > d_{THR} \end{cases}$ |

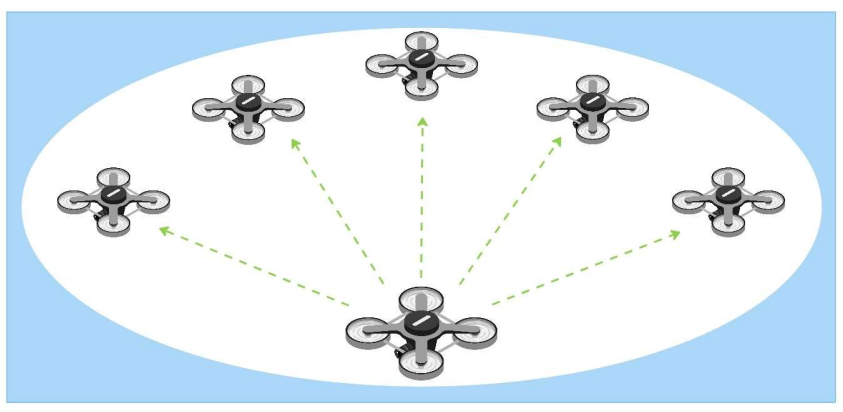

Fig.3: IR-UWB ranging data acquisition process from one UAV to the rest of the UAVs in the swarm

According to test results published by the authors in [28], the average energy consumption of the commercially available DW1000 IR-UWB module manufactured by Decawave and operating in Double-Sided Two-Way-Ranging (DS-TWR) mode is 3.55 mJ per one ranging, disregarding the time spent in the idle mode. Taking into account the average time of flight equal to 25 min and an average battery size of 5000 mAh at 14.8 V producing 74000 mWh, we can calculate the additional energy consumption posed by the IR-UWB measurements. By defining the IR-UWB measurement's frequency at 2 Hz we can estimate the power consumption per second, which would be equal to 7.1 mW. Thus, the overall energy consumed during the 25 min flight would be equal to $7.1\ mW \cdot 25/60\ h = 3\ mWh$, which is an equivalent of the flight time reduction of 0.12 second.

The ranging results taken between the UAVs twice a second for 25 minutes would accumulate 3000 ranging results that have to be transmitted to the GCS for further processing. Considering that the size of a message containing one ranging result along with the information about it could not exceed 50 bytes and an average data rate of a Sub-GHz telemetry radio module [29] is at around 40 Kbps with its corresponding transmitting current of 55 mA at 3 V, we can estimate the amount of energy required to transmit those messages to the GCS. Thus, the power consumption per second required to transmit the messages would be equal to $3\ V \cdot 55\ mA = 165\ mW$. The time required to send three thousand 50 byte messages at 40 Kbps would be equal to 30 seconds. Therefore, the overall energy consumed to transmit the messages containing IR-UWB ranging results over the telemetry radios from one of the UAVs in the swarm to the GCS would be equal to $165\ mW \cdot 0.5/60\ h = 1.375\ mWh$, which translates to the flight time reduction of 0.06 s.

Combining both, energy constraints posed by the IR-UWB measurements and radio transmission to the GCS, we are looking at the reduction in flight time of less than 1 second per drone per flight, which is not a significant amount of time for a UAV swarm operation. However, the real world energy consumption is usually higher than the calculations for ideal conditions. Although that needs to be further examined by executing real world tests.

To render our detection mechanism ineffective, an attacker would need to simultaneously spoof each GPS receiver of every drone in the swarm with an individual GPS signal, with respect to the real-time distance between them, which is virtually impossible and very resource demanding.

### *B. Detection mechanism integration*

The proposed spoofing attack detection mechanism is well suited for small and average size UAV swarms of up to 20 UAVs. It can be easily implemented in the form of software on the GCS's side. The GCS gets regular updates about the UAV's GPS position and upon request can also receive the IR-UWB or any other ranging data from the UAVs with the same frequency at any time.

Algorithm 1 implementing our GPS spoofing detection mechanism based on the criteria represented in Table II is given in the Appendix.

## V. Conclusion and Future work

GNSS and GPS spoofing attacks in particular, are a serious threat to the UAV swarm operation. As long as the GPS spoofing attack remains undetected, the attacked UAV is navigated by the attacker, which could lead to significant harm to the people around and result in a crash or UAV's theft. In this paper, we have studied the issue of the GPS spoofing attack detection and proposed a GPS spoofing detection mechanism for small and average size UAV swarms. The advantage of the proposed mechanism is that it does not require significant computational or energy resources, installment of additional antennas or modification of the existing hardware, and therefore is suitable even for micro UAVs. The proposed mechanism utilizes IR-UWB ranging, to confirm the GPS localization's plausibility, but is not limited to it and could be implemented using other similar ranging technologies. Another advantage of this detection mechanism is that it is very straightforward, and does not require a substantial data analysis, which is especially crucial for real-time UAV swarm operation. It should also be mentioned that this is the initial stage of this study and further implementation into a UAV swarm will follow in the near future. Performance, detection and false positive rates will be presented in the next stages of this research.

## Acknowledgment

This research has been partially funded by the Federal Ministry of Education and Research of Germany under grant numbers 16ES1131 and 16ES1128K.

## Appendix

**Algorithm 1.** GPS spoofing detection mechanism

```
import swarm_positions
import swarm_ranging
import math
class GPS_Distance:
```

```
  def __init__(self, timestamp: int, latitude: float,
                    longitude: float, altitude: float)
    self.timestamp: timestamp
    self.latitude = latitude
    self.longitude: = longitude
    self.altitude: = altitude

  def calc_ditance(self, other: "swarm_positions")
    distances = {}
    timestamp = min(self.timestamp, other.timestamp)
    distances["gps_timestamp"] = timestamp
    #Conversion from degrees to radians
    rad_lat_self = self.latitude * math.pi/180
    rad_lon_self = self.longitude * math.pi/180
    rad_lat_other = other.latitude * math.pi/180
    rad_lon_other = other.longitude * math.pi/180
    rad_delta = rad_lon_other - rad_lon_self
    R = 6371000 #earth radius in meters
    #Spherical law of Cosines
    d_flat = math.acos(math.sin(rad_lat_self)*
    math.sin(rad_lat_other)+math.cos(rad_lat_self)*
    math.cos(rad_lat_other)*math.cos(rad_delta)*R)
    #Euclidean distance with respect to altitude
    euclidean_distance = math.sqrt(d_flat**2 +
    (self.altitude - other.altitude)**2)
    distances["gps_dist"] = euclidian_distance
    return distances

class Spoofing_Detection:
  def __init__(self, gps_dist, uwb_dist,
               gps_timestamp, ubw_timestamp,
               time_threshold, dist_threshold)
    self.gps_dist = gps_dist
    self.uwb_dist = uwb_dist
    self.gps_timestamp = gps_timestamp
    self.uwb_timestamp = uwb_timestamp
    self.time_thr = time_threshold
    self.dist_thr = dist_threshold

  def distance_comparison(self):
    if math.isclose(0, self.gps_dist, abs_tol=10**-1):
      #if the GPS distance is less than 10cm
      Spoofing = true
      return Spoofing
    elif abs(self.uwb_dist - self.gps_dist)>self.dist_thr
     and abs(self.gps_timestamp - self.uwb_timestamp) >
            self.time_thr:
      Spoofing = true
      return Spoofing
    else:
      Spoofing = false
      return Spoofing

def main():
 fail_safe = false
 get_gps_dist = GPS_Distance.calc_ditance(UAV_1,UAV_i)
 Spoofing =
  Spoofing_Detection.distance_comparison(GPS,UWB)
   if Spooting == true:
      fail_safe = true

if __name__ == "__main__":
   main()
```